\documentclass[usenatbib]{mnras}
\usepackage[utf8]{inputenc}
\usepackage{amsmath}
\usepackage{amsfonts}
\usepackage{amssymb}
\usepackage{graphicx}
\usepackage{multirow}
\usepackage{hyperref}
\usepackage{booktabs}
\usepackage{threeparttable, tablefootnote}

\title[SN photometric classification and photo-z estimation]{Photometric classification and redshift estimation of LSST Supernovae}

\author[Dai et al.]{Mi Dai,$^{1}$\thanks{Email: mdai@physics.rutgers.edu} Steve Kuhlmann,$^{2}$ Yun Wang,$^{3}$ and Eve Kovacs$^{2}$
\vspace*{4pt}\\
       $^{1}$Department of Physics \& Astronomy, Rutgers, The State University of New Jersey, 136 Frelinghuysen Rd, Piscataway, NJ 08854, USA\\
       $^{2}$Argonne National Laboratory, 9700 South Cass Avenue, Lemont, IL 60439, USA\\
       $^{3}$IPAC, California Institute of Technology, Mail Code 314-6, 1200 East California Boulevard, Pasadena, CA 91125, USA}

\begin{document}

\maketitle

\begin{abstract}
Supernova (SN) classification and redshift estimation using photometric data only have become very important for the Large Synoptic Survey Telescope (LSST), given the large number of SNe that LSST will observe and the impossibility of spectroscopically following up all the SNe. We investigate the performance of a SN classifier that uses SN colors to classify LSST SNe with the Random Forest classification algorithm. Our classifier results in an AUC of 0.98 which represents excellent classification. We are able to obtain a photometric SN sample containing 99\% SNe Ia by choosing a probability threshold. We estimate the photometric redshifts (photo-z) of SNe in our sample by fitting the SN light curves using the SALT2 model with nested sampling. We obtain a mean bias ($\left<z_\mathrm{phot}-z_\mathrm{spec}\right>$) of 0.012 with $\sigma\left( \frac{z_\mathrm{phot}-z_\mathrm{spec}}{1+z_\mathrm{spec}}\right) = 0.0294$ without using a host-galaxy photo-z prior, and a mean bias ($\left<z_\mathrm{phot}-z_\mathrm{spec}\right>$) of 0.0017 with $\sigma\left( \frac{z_\mathrm{phot}-z_\mathrm{spec}}{1+z_\mathrm{spec}}\right) = 0.0116$ using a host-galaxy photo-z prior. 
Assuming a flat $\Lambda CDM$ model with $\Omega_m=0.3$, we obtain $\Omega_m$ of $0.305\pm0.008$ (statistical errors only), using the simulated LSST sample of photometric SNe Ia (with intrinsic scatter $\sigma_\mathrm{int}=0.11$) derived using our methodology without using host-galaxy photo-z prior.
Our method will help boost the power of SNe from the LSST as cosmological probes.
\end{abstract}


\section{Introduction}
Since the accelerating expansion of the Universe was discovered by observing distant Type Ia supernovae (SNe Ia) \citep{Riess98, Perlmutter99}, SNe Ia have been playing an important role in constraining the unknown cause behind the observed cosmic acceleration, or what we refer to as dark energy. The SNe we observe need to be correctly typed and accurate redshift information needs to be obtained, before the SNe Ia can be used to constrain cosmological models. With a sample size of $<1000$, it is possible to obtain correct types and redshifts of SNe via spectroscopy. Ongoing and planned surveys such as the Dark Energy Survey (DES) \citep{Bernstein2012}, and the Large Synoptic Survey Telescope (LSST) \citep{lsst}, will observe a dramatically increased number of SNe, making it difficult to spectroscopically follow up all the SNe with limited resources. SN cosmology will rely on photometric typing and redshift estimation. It is important to derive methods for reliable and accurate typing and redshift estimation using the photometric data of the SNe only. 

Current methods for SN classification include comparing the SN light curves against a set of SN templates (PSNID, \cite{sako2008}), or making a series of cuts based on the fitted results of certain SN Ia models \citep{bazin2011}, etc. \cite{kessler10spcc} describes and compares a list of methods that participated in the Supernova Photometric Classification Challenge (SPCC). Most recently, more efforts have been put in developing classification methods using machine learning techniques \citep{Lochner2016, Moller2016}. It is therefore important to explore which features work well in a machine learning algorithm. 

In this paper, we use realistic LSST SN simulations to study the performance of SN classification with the Random Forest classification algorithm, using SN colors as features for the first time, together with parameters from a general, model-independent function fit of the light curves.
Features used in machine-learning algorithms are typically derived from the properties of fits to a light-curve model, such as SALT2 \citep{guy2007,guy2010}. However, the SALT2-like models have been tuned on SN Ia data, and therefore have encoded correlations between parameters such as stretch and color, which are not correct for core-collapse (CC) SNe. This could lead to biased features for the CC SNe and make it more difficult for machine learning algorithms to type them correctly. Therefore we investigate the use of a general function (which is independent of a specific SN model) to fit the light curves.
Without utilizing the redshift information from SN host galaxies, we are able to obtain a photometrically classified SN Ia sample that is 99\% pure. We derive the photometric redshifts of the SNe by fitting their light curves using the SALT2 model with a nested sampling algorithm, and study the performance of this photometrically classified sample with photometric redshifts in constraining cosmology. We describe our simulations in Section \ref{sec:simulations}, our classification method in Section \ref{sec:classification}, our photo-z estimator in Section \ref{sec:photo-z}. The construction of a photometric Hubble Diagram is described in Section \ref{sec:cosmology}, followed by a summary and discussion in Section \ref{sec:summary}. 

\section{SN simulations}\label{sec:simulations}
We use the SNANA\footnote{http://snana.uchicago.edu/} \citep{Kessler09} software to generate realistic SN light curve simulations; it provides a simulation library that contains the survey condition information, and the LSST filter transmissions. 
The simulation library is created based on our current knowledge of LSST. 
Our simulation is based on 10 years of operation on 10 deep drilling fields of LSST. The actual number and location of the deep drilling fields, as well as the observing cadence, are still under study. The observing cadence will affect the final number of SNe that we can observe and the quality of the light curves. We generate mixed-type SNe by using SN rates for both Type Ia and non-Ia (CC) SNe. The SN rate is in the form of $\alpha(1+z)^\beta$. For SNe Ia, we use the rate measured by \cite{Dilday2008}, with $\alpha_\mathrm{Ia} = 2.6 \times 10^{-5} \, \mathrm{SNe}\, \mathrm{Mpc}^{-3}\, h^3_{70}\, \mathrm{yr}^{-1}$ and $\beta_\mathrm{Ia} = 1.5$; for the 
CC SNe, we assume $\alpha_\mathrm{CC} = 6.8 \times 10^{-5} \, \mathrm{SNe}\, \mathrm{Mpc}^{-3}\, h^3_{70}\, \mathrm{yr}^{-1}$ and $\beta_\mathrm{CC} = 3.6$, following \cite{Bernstein2012,lsst}. The Ia light curves are generated using the extended SALT2 model, provided by SNANA, which extends the standard SALT2 model \citep{guy2007,guy2010} on both ends of the spectral template
via extrapolation (with the extended wavelength range from 300 $\mathrm\AA$ to 18000 $\mathrm\AA$).
The use of this extended model is essential in generating light curves in all six LSST bands, since with the standard SALT2 model some bands with certain redshifts will not be generated if they fall out of the wavelength range of the standard model. However, since there is either little flux (for the UV bands) or the flux is very noisy (for the IR bands) outside of the wavelength range of the standard SALT2 model, the light curves generated in those bands are basically constant noise. Section \ref{subsec:SALT2} gives a brief review of the SALT2 model. 
A color dispersion is applied by SNANA based on the color dispersion derived by \cite{guy2010}. The CC light curves are generated through a set of spectral templates, provided in the SNANA software. (For more details, see \cite{kessler10spcc}.)
We make a quality cut during the simulation by requiring that the photometry in at least 3 bands have a maximum signal-to-noise-ratio (SNR) greater than 5. A total of 144246 SNe are obtained, including 62147 Ia, 67631 II, and 14468 Ibc. The redshift range of our simulation is from 0.01 to 1.2. We also apply other quality cuts on this sample in different stages, which are discussed in section \ref{subsec:quality cuts}.

\section{SN classification}\label{sec:classification}

\subsection{Using SN colors for classification}\label{subsec:cla using color}
In \cite{Wang2015}, SN colors are used to build an analytic photo-z estimator and are shown to have very good performance. Inspired by the ability of using SN colors only to estimate the SN photo-z, we expand the usage of SN colors in SN classification. The SN light curves are first fitted into a general functional form that is described in detail in Section \ref{subsec:bazinfit}, so that a relatively accurate peak magnitude for each band can be obtained. The peak magnitudes (converted from peak fluxes) are then calculated using the fitted parameters (Eq. \ref{eq:bazintmax} and \ref{eq:bazinfmax}), and the colors are calculated as the difference between the peak magnitudes of the two adjacent bands. 

\begin{equation}\label{eq:color}
c_{ij} = m_{p, i} - m_{p, j}
\end{equation}
where $i$ and $j$ represents two of the adjacent bands of the LSST filters (ugrizY).

We find that the SN colors can be used in SN classification with very good performance using the machine learning classification algorithms when they are used together with the two parameters from the general parametrization that describes the rising and falling time of the SN light curves.

\subsection{General parametrization of SN light curves}\label{subsec:bazinfit}
The SN light curves (both Type Ia and non-Ia) are found to be well fitted using a general functional form; it has no specific physical motivation, but describes the light curve shapes in a model-independent manner. Following \cite{Bazin09}, we use the following equation to fit the SN light curves (we refer to this as the ``Bazin function"): 

\begin{equation}\label{eq:bazinfunc} 
f(t) = A\frac{\exp^{-(t-t_0)/t_{\mathrm{fall}}}}{1+\exp^{-(t-t_0)/t_{\mathrm{rise}}}} + B
\end{equation}

In this equation, $t_\mathrm{fall}$ and $t_\mathrm{rise}$ measure the declining and rising time of the light curve, $A$ is the normalization constant, and $B$ is a constant term. We should note that $t_0$ is not exactly the date of maximum flux (date at the peak of the light curve), and similarly $A$ is not exactly the maximum flux. By calculating the derivative of the function we are able to obtain the functional form of date at the maximum flux and the value of the maximum flux:

\begin{equation}\label{eq:bazintmax}
t_{\mathrm{max}} = t_0 + t_{\mathrm{rise}}\ln(\frac{t_{\mathrm{fall}}}{t_{\mathrm{rise}}}-1)
\end{equation}
\begin{equation}\label{eq:bazinfmax}
f(t_{\mathrm{max}}) = A x^x(1-x)^{1-x} +B, 
x=\frac{t_{\mathrm{rise}}}{t_{\mathrm{fall}}}
\end{equation}

The equation above also indicates that we should set $\frac{t_{\mathrm{rise}}}{t_{\mathrm{fall}}} <1$ in order to have a meaningful $t_\mathrm{max}$. This constraint is useful in excluding some of the bad fits. More details on the quality cuts are described in Section \ref{subsec:quality cuts}.

For higher redshift SNe, there is little flux in the u, g band. At the other end of the spectra, the Y band data are usually noisy.
We notice that the Bazin function does not fit well for these bands with low SNR. So we fit the light curve with different forms depending on the SNR of the band being fitted. For $\textrm{SNR} > 5$, the light curve is fitted using Eq. \ref{eq:bazinfunc}, otherwise the light curve is fitted to a constant $f(t) = B$.

We utilize the curvefit procedure in python scipy\footnote{https://www.scipy.org/}. It is necessary to set initial values and limits for the parameters being fitted. To achieve better results, we do the fit in two steps with different initial conditions and parameter limits. We list the initial values and parameter limits in Table \ref{tb:initial_condition}. The initial values of the 2nd fit are calculated using the results of the 1st fit. We define a ``successful fit" as satisfying the following conditions: $t_{\mathrm{fall}} > t_{\mathrm{rise}} > 1$. The successfully fitted bands in the 1st fit are kept and the median of the parameters of such bands are used as the initial value in the 2nd fit. 
A 2nd fit is also performed when the constant fitting returns a $B$ value larger than 5; this usually happens for the Y band where the SNR is lower than 5 but there is indeed signal and can be fitted with Eq. \ref{eq:bazinfunc}.

The peak magnitude in each band is calculated as: 
\begin{equation}
m_p = -2.5 \log_{10}(f_\mathrm{max}) + \mathrm{zero{\hbox{-}}point}
\end{equation}
where $f_\mathrm{max}$ is calculated using Eq. \ref{eq:bazinfmax}. For any band that is fitted to a constant, $t_\mathrm{fall}$, $t_\mathrm{rise}$, and $m_p$ are set to 0, in which case an error-weighted color is obtained. The colors are calculated using Eq. \ref{eq:color}, regardless of the value of $m_p$. We notice that the colors can be 0, or the opposite of the peak magnitude in one band, instead of actual colors, when the zero peak magnitude is used in calculating the colors. We treat this as a property of our sample and pass it to the classifier.

Since the Bazin function is only a parametrized function to describe the SN light curve shapes in general, we do not expect it to deliver as accurate Ia peak magnitudes as a Ia-specific model such as SALT2, which is trained on a set of well-sampled Ia photometry and spectra. While the SALT2 model fits the multi-band SN light curves simultaneously, imprinting a known color relation (from training) into the resultant peak magnitudes, our approach is to obtain single band peak magnitudes independently, without knowing such relations, so that the peak colors are not biased toward the known model -- this is necessary for obtaining the CC SN peak colors which may be different from the assumed Ia color model. 
To justify our claims, we have compared the peak magnitudes obtained from both the Bazin fit and the SALT2 fit to the true peak magnitudes calculated from the simulation. A detailed discussion can be found in Appendix \ref{appx2}.

\begin{table*}
\centering
\caption{2-step general parametrization fit: initial conditions and parameter limits}
\label{tb:initial_condition}
\begin{tabular}{|l|l|l|l|l|}
\hline
 & \multicolumn{2}{l|}{1st step} & \multicolumn{2}{l|}{2nd step} \\ \hline
 &    initial value  &  limits   &  initial value   &  limits    \\ \hline
$A$ &  flux at peak  & [0, inf]  & flux at peak  & [0, inf] \\ \hline
$t_0$ &  time at peak & [-inf, inf]  & median($t_0$) & fixed     \\ \hline
$t_{\mathrm{fall}}$ &  15  & [0, inf]  & median($t_{\mathrm{fall}}$)  & [1, inf]  \\ \hline
$t_{\mathrm{fall}}$  & 5   & [0, inf]  & median($t_{\mathrm{rise}}$)  & [1, inf]  \\ \hline
$B$ & 0   & [-inf, inf]  &  0   & [-inf, inf] \\ \hline
\end{tabular}
\end{table*}

\subsection{Quality cuts}\label{subsec:quality cuts}
In order to obtain a high quality sample, we apply several quality cuts, both before and after the general function fit. Before the fitting, we require that the max SNR is greater than 5 for at least 3 bands, and that at least 3 bands have 1 point before the peak and 2 points after the peak, including at least one of the $\textrm{SNR}>5$ bands. These cuts ensure that the light curve has a well-defined peak and not too noisy in at least one band so that at least one successful fit is achieved in the first step described above. 

After all the pre-fitting quality cuts are applied, the fitting program is able to return a set of parameter values for most of the SNe, although some of the values are not in a reasonable range. So we make a series of cuts based on the parameter distributions. The cuts we used are listed below:

\begin{itemize}
\item $t_{\mathrm{rise}} > 1$, and $t_{\mathrm{rise}}$ not close to 1 with tolerance = 0.01
\item $-20<B<20$
\item $\chi^2/d.o.f < 10$
\item $t_{\mathrm{fall}} < 150$
\item $t_{\mathrm{rise}} < t_{\mathrm{fall}}$ 
\item $A < 5000$
\item $A_{\mathrm{err}} < 100$
\item $t_{0,\mathrm{err}} < 50$
\item $t_{\mathrm{fall,err}} < 100$
\item $t_{\mathrm{rise,err}} < 50$
\item $A(Y), A(u) < 1000$
\end{itemize}

These cuts also serve to exclude some of the non-Ia's, especially type II's which have a rather larger $t_{\mathrm{fall}}$ value. The remaining fractions after cuts for each of the three SN types in the simulations are Ia, Ibc and II are 55\%, 38\% and 15\%, respectively.

This after-cut sample is used to test our classification algorithm, which contains 68\% Ia's, 11\% Ibc's and 20\% II's. 
We notice that the $\chi^2$ cut eliminates almost all of the low-z ($z<0.3$) SNe, since the low-z SNe usually have very high SNR and have a second peak in the redder bands, which cannot be well-fitted using Eq. \ref{eq:bazinfunc}, and thus result in very large $\chi^2$ per d.o.f.

Detailed lists of the remaining number of SNe after each cut are shown in Appendix \ref{appx1}.

\subsection{SN classification with random forest algorithm}\label{subsec:ml concepts}

Machine learning algorithms are used in SN classification recently \citep{Lochner2016, Moller2016}, and have excellent performance when the features used in the classifier are carefully selected. Here we choose the Random Forest algorithm to demonstrate the performance of classification using SN colors. We adopt a code similar to the one used in host galaxy identification by \cite{gupta2016}, and modified it to suit our needs. For details about the algorithm, see \cite{randomforest}. There are also many other machine learning algorithms that can be used, most of which are very easy to implement. The comparison of performance for several commonly used machine learning algorithms can be found in \cite{Lochner2016}.

As described in Section \ref{subsec:cla using color}, a total number of 17 features are passed to our classifier. The features are: 12 Bazin-fit parameters $t_\mathrm{fall}(u)$, $t_\mathrm{rise}(u)$, $t_\mathrm{fall}(g)$, $t_\mathrm{rise}(g)$, $t_\mathrm{fall}(r)$, $t_\mathrm{rise}(r)$, $t_\mathrm{fall}(i)$, $t_\mathrm{rise}(i)$, $t_\mathrm{fall}(z)$, $t_\mathrm{rise}(z)$, $t_\mathrm{fall}(Y)$, $t_\mathrm{rise}(Y)$, and 5 colors $c_{ug}$, $c_{gr}$, $c_{ri}$, $c_{iz}$, $c_{zY}$. 

We now summarize the concepts that are commonly used in presenting the classification results.

\subsubsection{Confusion matrix}

For a binary classification problem, a confusion matrix is defined in Table \ref{confusion_m}. In our case, the two classes are `Ia' (Yes) and `non-Ia' (No). 

\begin{table*}
\centering
\caption{Confusion matrix for a binary classification}
\label{confusion_m}
\begin{tabular}{|c|c|c|c|}
\hline
\multicolumn{2}{|c|}{\multirow{2}{*}{}} & \multicolumn{2}{c|}{Predicted Class} \\ \cline{3-4} 
\multicolumn{2}{|c|}{}              &   Yes      &    No      \\ \hline
\multirow{2}{*}{Actual Class}       &   Yes     & Ture Positive (TP) & False Negative (FN) \\ \cline{2-4} 
                            &   No     & False Positive (FP) & True Negative (TN) \\ \hline
\end{tabular}
\end{table*}

\subsubsection{Receiver operating characteristic (ROC) curves}

We define the true positive rate (TPR) and the false positive rate (FPR) as the following (according to the confusion matrix):

\begin{equation}
\mathrm{TPR} = \frac{\mathrm{TP}}{\mathrm{TP}+\mathrm{FN}}
\end{equation}
\begin{equation}
\mathrm{FPR} = \frac{\mathrm{FP}}{\mathrm{FP}+\mathrm{TN}}
\end{equation}

By varying the probability threshold within a classifier in determining the class, different values of TPR and FPR are returned. The ROC curve is defined as TPR vs FPR, since we would expect an excellent classifier to have high TPR with low FPR. Another value that is often used in comparing classification results is the area-under-the-curve (AUC) of a ROC curve. For perfect classification, $\mathrm{AUC}=1$, the ROC curve behaves as a step function, while for a random classification, $\mathrm{AUC}=0.5$, the ROC curve behaves as a diagonal line. An AUC that is larger than 0.9 usually represents excellent classification.

\subsubsection{Efficiency and Purity}

We can also define the efficiency and purity using the confusion matrix:
\begin{equation}
\mathrm{efficiency} = \frac{\mathrm{TP}}{\mathrm{TP}+\mathrm{FN}}
\end{equation}
\begin{equation}
\mathrm{purity} = \frac{\mathrm{TP}}{\mathrm{TP}+\mathrm{FP}}
\end{equation}

To achieve higher purity usually means sacrificing the efficiency, and vice versa, given that the probability threshold is varied. 

\subsection{Training sample size determination}
The classification algorithms rely on a training sample with known types to predict types for the test sample. There are generally two ways that a training sample can be obtained: one is to use a spectroscopic sample from the same survey, but the sample size can be relatively small, compared to the large number of SNe that LSST can observe; the other is simply using realistic simulations, so the sample size can be as large as we need, representing good statistics of the test sample.

We compare the effects on the classification results by varying the training sample size as a fraction of 0.05, 0.1, 0.3, 0.5 and 0.9 of the total sample. We find that a training set of 0.05 or 0.1 fraction of the whole sample results in an AUC of 0.96, while a fraction greater or equal to 0.3 results in an AUC of 0.98, which indicates that a large enough training sample size is required to best represent the sample and thus leads to better classification performance. However, as our sample size after all the quality cuts is $\sim 39000$ including all types, a 0.1 fraction with $\sim 3900$ SNe is already larger than current spectroscopically confirmed data sets. This means that a spectroscopic training set for classification will be challenging to obtain. We also conclude that a training sample with comparable size to the test sample will result in best performance. In this paper, we show the classification result with the 0.3-fraction training sample. We use the same sample for the following analysis, in order to obtain as large a sample size as we can for the cosmological analysis. When dealing with real data in the future, a simulated sample with the same size as the real sample can be used for best performance.

\subsection{Classification results}

\begin{figure}
\centering
\includegraphics[width=0.4\textwidth]{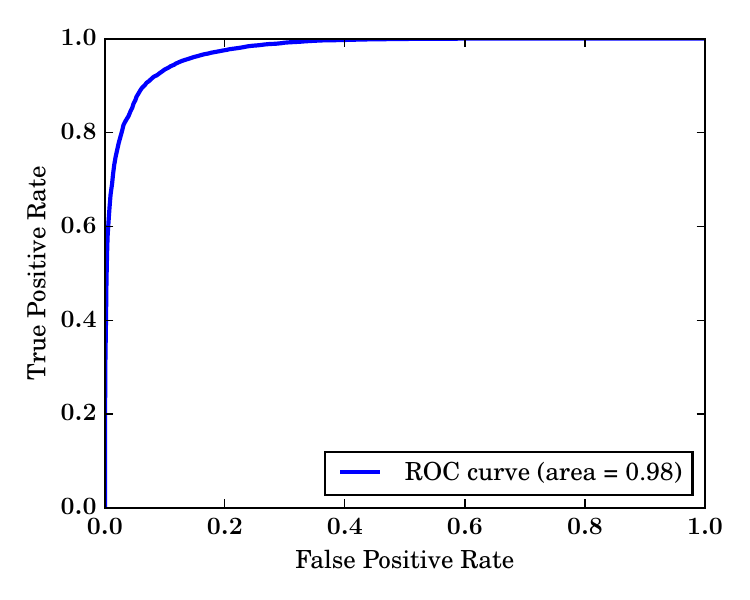}
\caption{ROC curve for our classification result.}
\label{fig:roc}
\end{figure}

We now present the classification results using the concepts defined in Section \ref{subsec:ml concepts}. Fig. \ref{fig:roc} shows the ROC curve for our classification, with an AUC of 0.98, indicating that we reach excellent classification by using the features we described in Section \ref{subsec:cla using color}. This AUC value is comparable to recent studies \citep{Lochner2016, Moller2016} with different data sample or simulations.

\begin{figure}
\centering
\includegraphics[width=0.4\textwidth]{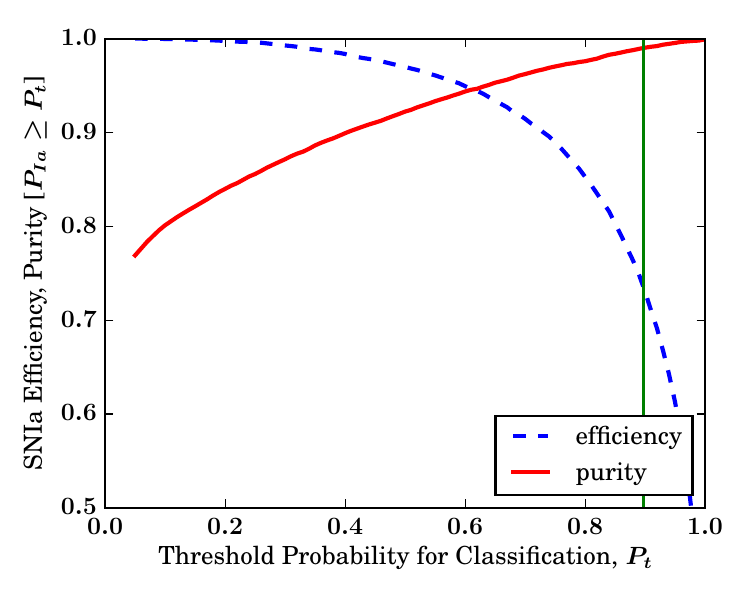}
\caption{Purity and efficiency curves for our classification result, red solid line shows the purity curve with respect to the threshold probability chosen, blue dashed curve shows the efficiency curve. 
The vertical line indicates the threshold probability for 99\% purity.}
\label{fig:eff_pur}
\end{figure}

Fig. \ref{fig:eff_pur} shows the purity and efficiency curves as the threshold probability varies. High purities can be obtained by sacrificing some efficiency. We notice that a 90\% - 95\% purity can be easily achieved with efficiency larger than 90\%. While we aim at a purity of 99\% for the following cosmological analysis, the efficiency dropped to 74\%. Note that this efficiency is the classification only efficiency, not including the quality cuts through all the procedures. Our analysis results in a photometric sample with 13744 SNe with 99\% purity. 

\section{SN photometric redshift}\label{sec:photo-z}

Accurate redshift information is essential in constructing the Hubble Diagram and constraining cosmology. Analyses of past and ongoing surveys rely on spectroscopic redshifts -- either from the SN spectra or the host-galaxy spectra. With LSST, it is impractical for us to obtain spectroscopic redshifts for all the SNe; it is also unclear whether we will be able to obtain spectroscopic redshifts for the host galaxies of all the SNe; thus it is useful to develop methods for SN redshift estimations using the photometric data. Currently two kinds of approaches are proposed for SN photo-z estimation: one is analytic utilizing the multi-band SN colors \citep{wang2007photoz,wang2007photoz2,Wang2015}, the other is template-based by fitting the light curves into SN Ia models  \citep{kessler10photoz,palanque10photoz}. We adopt the template approach in this paper, by fitting the SN light curves using the commonly used SALT2 model, using the nested-sampling method. Our fits are performed using the SNCosmo\footnote{https://sncosmo.readthedocs.io/} package. We also investigate the performance of our photo-z estimator using a host-galaxy photo-z prior.

\subsection{The SALT2 model}\label{subsec:SALT2}

The SALT2 model \citep{guy2007,guy2010} provides an average spectral sequence and its higher order variations, as well as a color variation law, which can be used to fit SN Ia light curves using several parameters. The model flux is given as:
\begin{equation}
\label{eq:model}
\frac{dF}{d\lambda}(p,\lambda) = x_0 \times [M_0(p,\lambda)+x_1M_1(p,\lambda)+\ldots] \times \exp[c\,CL(\lambda)]
\end{equation} 
where $p$ is the phase (rest-frame time from the maximum light), $\lambda$ is the rest-frame wavelength, $M_0$ and $M_1$ are the spectral sequence and its first order variation, $CL$ is the color law. $x_0$, $x_1$ and $c$ are light curve parameters that describe the amplitude, stretch and color of the light curve; they are fitted in a fitting process in which the date of maximum $t_0$ and the redshift $z$ can also be fitted simultaneously.
  
In this paper, we use an extended model that covers wider wavelength ranges (from 300 $\mathrm\AA$ to 18000 $\mathrm\AA$) than the standard SALT2 model.

\subsection{2-stage fit using nested sampling}\label{subsec:SALT2 fit}

We choose to use the nested sampling method for light curve fitting using SALT2. We find that nested sampling results in better photo-z estimates, compared to the normal maximum likelihood method using {\it MINUIT}. Using the same fitting procedures as we described in this section, but only changing the fitting method, we find that nested sampling results in a photo-z outlier ($\left|\frac{z_\mathrm{phot}-z_\mathrm{spec}}{1+z_\mathrm{spec}}\right|>0.1$) fraction of 1.6$\%$ before applying further cuts, while MINUIT results in a photo-z outlier fraction of 9.6$\%$. The difference in fitting performance could be due to the fact that MINUIT sometimes fails to find the true minima of parameters when photo-z is fitted simultaneously. Nested sampling is a more robust technique since, unlike MINUIT, it does not converge on a local minimum. 

The model is not well characterized in the UV region with a dramatically increased uncertainty, which can lead to a wrong fit if the parameter limits are not set correctly. 

We take advantage of a 2-stage fit again in the SALT2 fit. Now we describe the two stages in detail:

For the initial fit, we aim at locating the right ranges of the parameters. The model covariance is not used in this fit, only the statistical errors from the photometry are included. We set the parameter limits as follows: $0.01<z<1.2$, $|x_1|<5$, $|c|<0.5$, 
$t_\mathrm{min}-15<t_0<t_\mathrm{max}+15$; and the bound of the amplitude parameter, $x_0$, is determined internally by SNCosmo.

The second fit is limited to a smaller range determined by the 1st fit, with $x_1$, $c$, $t_0$ limited to a 3-sigma range from the mean value of the 1st fit, while $x_0$ bounds are still ``guessed" by {\it SNCosmo}. We limit the redshift to a rather larger range: $z_\mathrm{ini} \pm 10\sigma_z$, in order to better estimate the final uncertainty of $z$. We find that the $3\sigma$ limit for the other three parameters is necessary for obtaining a good fit, since the fit can easily be trapped in an unreasonable parameter region where the $\chi^2$ is very small due to a large value of  uncertainty in the UV bands. This fitting deficiency has been observed in another analysis \citep{Dai2016}, which uses Markov Chain Monte Carlo (MCMC) method to fit light curves to the SALT2 model. In \cite{Dai2016}, the model covariance is kept fixed to mimic and reproduce the original SALT2 result. When we are simultaneously fitting the redshift, fixing the covariance is not applicable. We also use another condition in the fitting that helps to reduce this fitting bias: for SNe with redshifts (from the initial fit) that are less than 0.65, all 6 bands are used in the fitting; for those that have redshifts greater than 0.65, the u and g bands are excluded from the fitting. The 0.65 line is determined by observing the fitted results using all 6 bands and determining where the bias starts to occur. A better characterized model in the UV band can be very useful in all aspects. 

By applying the 2-step fit described above, we obtain a set of photo-z's with an accuracy
$\sigma \left(\frac{z_\mathrm{phot}-z_\mathrm{spec}}{1+z_\mathrm{spec}}\right) = 0.0294$, and a mean bias ($\left<z_\mathrm{phot}-z_\mathrm{spec}\right>$) of 0.0120, after applying a cut on the reduced $\chi^2$ of the SALT2 fit ($\chi^2_\mathrm{red}<1.5$). The results are shown in Fig. \ref{fig:dz-over-1plusz-distr-nohost}. The outlier ($\left|\frac{z_\mathrm{phot}-z_\mathrm{spec}}{1+z_\mathrm{spec}}\right|>0.1$) fraction is 1.12\%. 
We also show that our method results in accurate photo-z errors, as illustrated in Fig. \ref{fig:photoz-distr}. In Fig. \ref{fig:photoz-distr}, the histograms of $(z_\mathrm{phot}-z_\mathrm{spec})/z_\mathrm{err}$ are plotted in different redshift ranges, where the $z_\mathrm{err}$ is the error in the photo-z which is output from the SALT2 fitting. The histograms are fitted with a Gaussian function. With the fitted $\sigma$ close to 1, we conclude that photo-z error estimation from the SALT2 fitting is accurate. Such a fit can only be achieved when the SALT2 model covariance is included in the fitting and the parameter limits are carefully chosen.

\subsection{Effect of host galaxy priors}

We investigate the effect of using a host-galaxy photo-z prior in the SALT2 fitting. We apply a Gaussian prior with mean and sigma values set as the host-galaxy photo-z and error from a simulated host-galaxy library for LSST. 
The SNe with host-galaxy photo-z smaller than 0.01 or greater than 1.2 are dropped.
{Using this host-galaxy photo-z prior, we obtain a set of photo-z's with an accuracy
$\sigma \left(\frac{z_\mathrm{phot}-z_\mathrm{spec}}{1+z_\mathrm{spec}}\right) = 0.0116$, and a mean bias ($\left<z_\mathrm{phot}-z_\mathrm{spec}\right>$) of 0.0017, after applying a cut on the reduced $\chi^2$ of the SALT2 fit ($\chi^2_\mathrm{red}<1.5$). The outlier ($\left|\frac{z_\mathrm{phot}-z_\mathrm{spec}}{1+z_\mathrm{spec}}\right|>0.1$) fraction is 0.16\%.
The results are shown in Fig. \ref{fig:dz-over-1plusz-distr-whost}. 
Using a host galaxy photo-z prior leads to significant improvement in the photo-z estimation, although the currently available LSST host galaxy library that we have used may have optimistic host galaxy photo-z errors. We will  re-evaluate the performance of our photo-z estimator with host-galaxy priors when a more realistic LSST host galaxy library becomes available.

\begin{figure}
\centering
\includegraphics[width=0.4\textwidth]{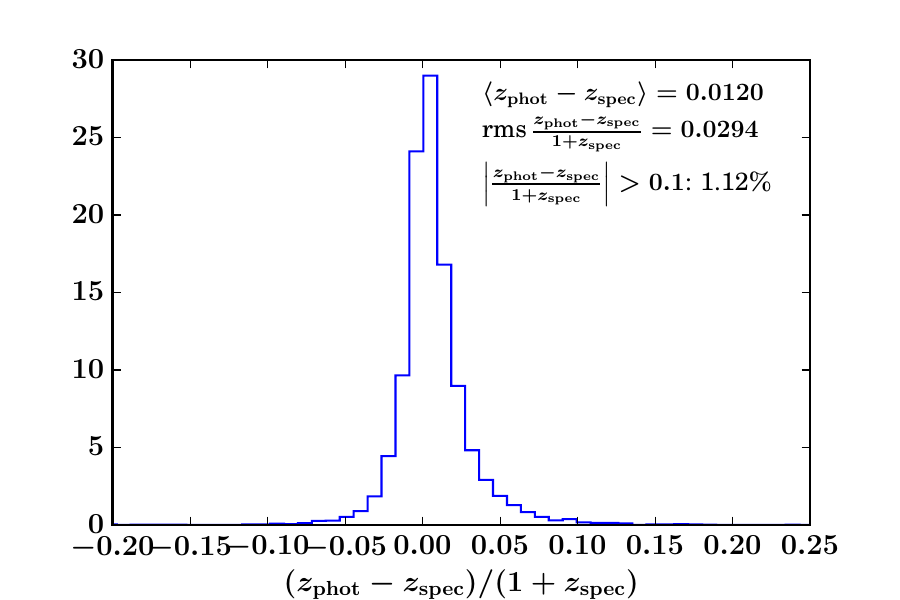}
\caption{Distribution of $(z_\mathrm{phot}-z_\mathrm{spec})/(1 + z_\mathrm{spec})$, with no host-galaxy photo-z prior}
\label{fig:dz-over-1plusz-distr-nohost}
\end{figure}

\begin{figure*}
\centering
\includegraphics[width=\textwidth]{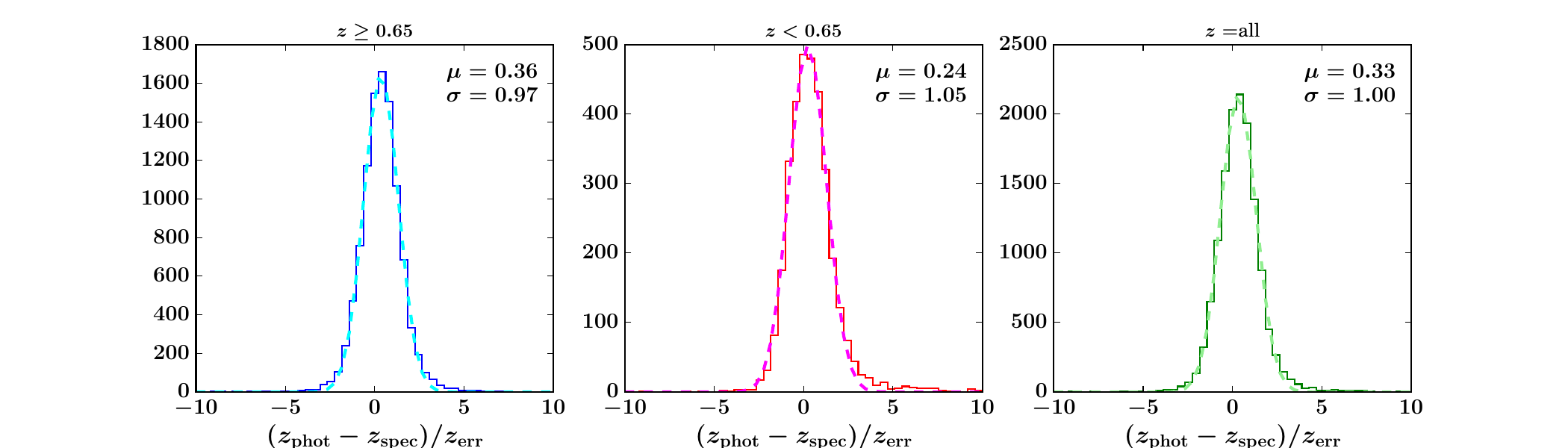}
\caption{Distribution of $(z_\mathrm{phot}-z_\mathrm{spec})/z_\mathrm{err}$ in different redshift ranges. Left: $z \geq 0.65$, middle: $z < 0.65$, right: z in the whole sample range. Dashed lines are from Gaussian fits with best-fit value shown in the right corner.}
\label{fig:photoz-distr}
\end{figure*}

\begin{figure}
\centering
\includegraphics[width=0.4\textwidth]{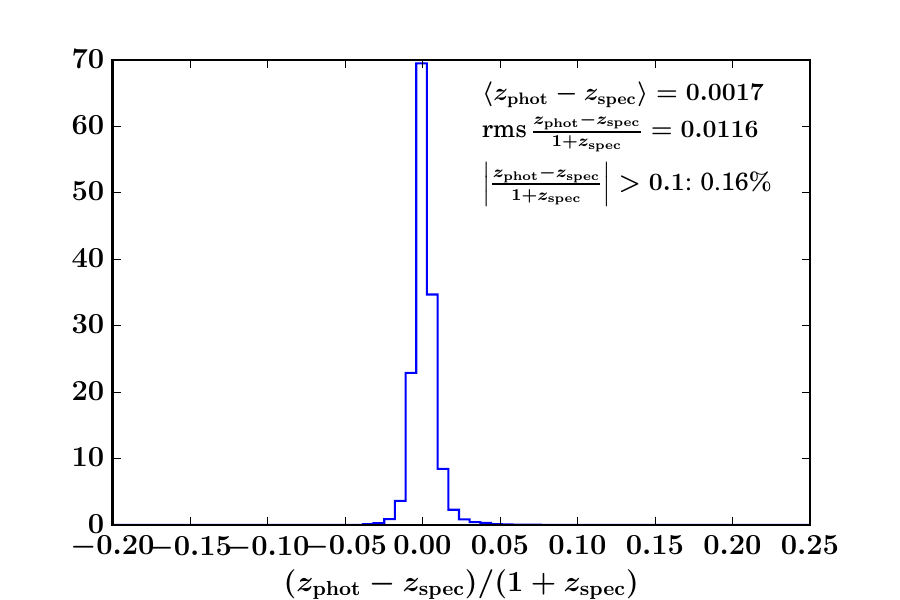}
\caption{Distribution of $(z_\mathrm{phot}-z_\mathrm{spec})/(1 + z_\mathrm{spec})$, with host-galaxy photo-z prior}
\label{fig:dz-over-1plusz-distr-whost}
\end{figure}

\section{Fitting cosmology}\label{sec:cosmology}

Our final step is to examine the performance of using our photometric-only SN Ia sample (with CC contamination) in constraining cosmology. We use the SALT2 parameters ($x_0$, $x_1$, $c$) obtained from Section \ref{subsec:SALT2 fit} to calculate the distance modulus $\mu$:
\begin{equation}
\label{eq:distmod}
\mu = m_B + \alpha \, x_1 - \beta \, c + \mathcal{M},
\end{equation}
where $\alpha$, $\beta$ and $\mathcal{M}$ are nuisance parameters, and $m_B=-2.5\log_{10}(x_0)$.

We have simulated the data assuming the $\Lambda \mathrm{CDM}$ model with $\Omega_m=0.3$ and  a flat Universe. The $\chi^2$ is calculated as:
\begin{equation}
\chi^2 = \sum_i \frac {(\mu_i-\mu_{\mathrm{model},i})^2}{\sigma^2_i}
\end{equation}

where
\begin{multline}
\sigma^2 = \sigma^2_{m_B} + \alpha^2 \sigma_{x_1}^2 + \beta^2 \sigma_{c}^2 + \left(\frac{\partial \mu_\mathrm{model}}{\partial z}\right)^2\sigma_z^2 + \sigma_\mathrm{int}^2 \\ 
+ 2\alpha \mathrm{Cov}_{m_B,x_1} - 2\beta \mathrm{Cov}_{m_B,c}  
-2 \left(\frac{\partial \mu_\mathrm{model}}{\partial z}\right)\mathrm{Cov}_{m_B,z}
-2\alpha\beta \mathrm{Cov}_{x_1,c} \\
-2 \alpha \left(\frac{\partial \mu_\mathrm{model}}{\partial z}\right)\mathrm{Cov}_{x_1,z}
+2 \beta \left(\frac{\partial \mu_\mathrm{model}}{\partial z}\right)\mathrm{Cov}_{c,z}
\end{multline}

The ${\partial \mu_\mathrm{model}}/{\partial z}$ term is calculated numerically by:
\begin{equation}
\frac{\partial \mu_\mathrm{model}}{\partial z} = \frac{5}{\log 10}\left(\frac{1}{1+z} + \frac{1}{r(z)}\frac{\partial r(z)}{\partial z}\right)
\end{equation}
where $r(z)$ is the comoving distance that depends on the given cosmological model.

We set $\sigma_\mathrm{int}$ to be a constant value and vary it with different values to see whether it affects the fitting outcome. Note that our simulation is generated using a more complicated intrinsic scatter model, so using this constant $\sigma_\mathrm{int}$ may introduce a bias. Only the statistical uncertainties are considered in this fit. The fit is performed using the CosmoMC software \citep{lewis2002}\footnote{http://cosmologist.info/cosmomc/}.

Before fitting to cosmological models, a bias correction term is calculated and applied to the distance modulus $\mu$ in Eq. \eqref{eq:distmod}. 
The bias correction term is
determined using a separate ``bias correction" set of simulation generated assuming a different cosmology (with $\Omega_m = 0.27$ and $w=-1$), with a sample size similar to our original data set. We apply our methodology for photometric classification and photo-z estimation to this separate data set, including the same cuts, and calculate the bias in 20 redshift bins by taking the mean of the differences between the fitted distance moduli, $\mu_\mathrm{fit}$, and the true distance moduli, $\mu_\mathrm{true}$ of the SNe in each bin:
\begin{equation}
\Delta\mu (z_i) = \left< \mu_\mathrm{fit} - \mu_\mathrm{true}\right>_{z_i}
\end{equation}
where $\mu_\mathrm{true}$ is calculated from the input cosmological model, and $\mu_\mathrm{fit}$ is calculated from the fitted SALT2 parameters, with $\alpha$ and $\beta$ set to be the simulated values.

The correction for each SN is then obtained using linear interpolation. So the bias-corrected distance of each SN becomes:
\begin{equation}
\mu_\mathrm{\scriptscriptstyle SN} = \mu_\mathrm{\scriptscriptstyle fit,SN} - \Delta\mu_\mathrm{\scriptscriptstyle SN} 
\end{equation}

\subsection{Ellipse cut and other quality cuts}

We utilize an ellipse cut (Fig. \ref{fig:ellipse_cut}) to exclude the SNe with extreme values in the $x_1$-$c$ plane.  We adopt a similar cut as in \cite{bazin2011} and \cite{campbell2013}. The ellipse we draw has semi-axes $a_\mathrm{x_1}=3$, and $a_\mathrm{c}=0.25$, centred at $(x_1,c)=(-0.2,0)$. Note that this cut excludes a higher fraction of non-Ia's than Ia's in our sample, and thus improves the purity of the final sample (99.7\%). 

We also require the photo-z to be greater than 0.2, since the low-z SNe are already excluded in the general-function-fit step before classification; we require the photo-z error to be less than 0.1.

Our final sample for cosmological analysis has a total number of 12618 SNe including 12586 (99.7\%) SNe Ia and 32 (0.3\%) CC SNe.
We show the marginalized means of the parameters in Table \ref{table:cosmo-par}. The Hubble Diagram is shown in Fig. \ref{fig:hd}. 

\begin{figure*}
\centering
\includegraphics[width=0.4\textwidth]{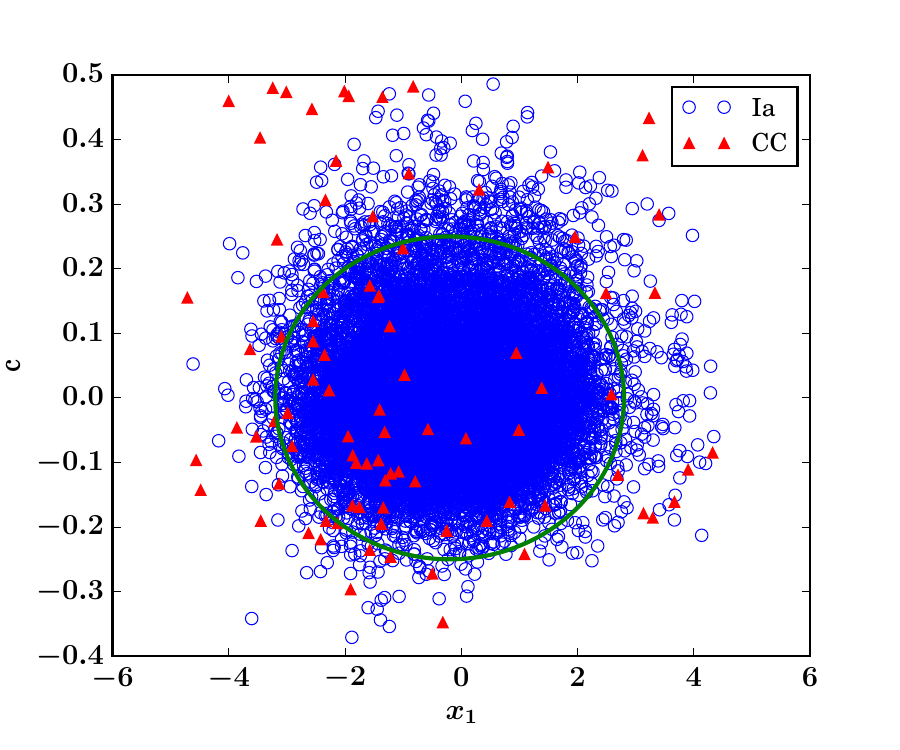}
\caption{Ellipse cut for SALT2 parameters $x_1$ and $c$}
\label{fig:ellipse_cut}
\end{figure*}

\begin{figure*}
\centering
\includegraphics[width=\textwidth]{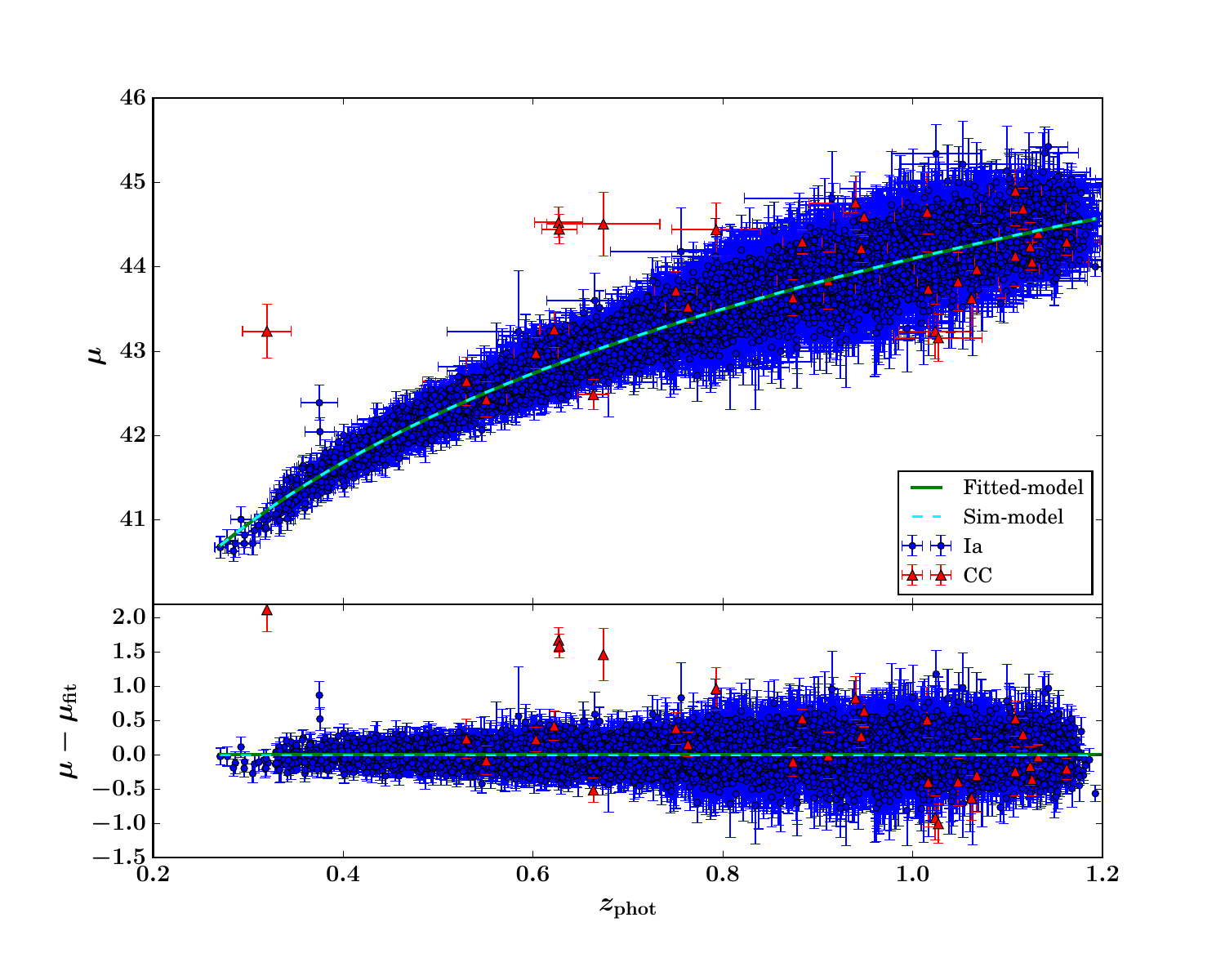}
\caption{Hubble Diagram of our photometric SN Ia sample derived without using host-galaxy photo-z prior. Blue dots are true SNe Ia, red triangles are Core-collapse SNe that are classified as Ia's. Green solid line is the  fitted cosmology and cyan dashed line is the simulated cosmology.}
\label{fig:hd}
\end{figure*}

\begin{table*}
\centering
\caption{Simulation input parameter values and the marginalized means of the cosmological parameters obtained using the photometric SN Ia sample derived without using host-galaxy photo-z prior.}
\label{table:cosmo-par}
\begin{tabular}{ccccccc}
\hline
 & $\Omega_m$ & $\alpha$ & $\beta$ & $\mathcal{M}$ & $\chi^2$ & d.o.f \\ \hline
Sim input & $0.30$ & $0.135$ & $3.19$ & $13.43$ & -  & - \\ \hline
$\sigma_\mathrm{int}=0.1$ & $0.299\pm0.008$ & $0.119\pm0.001$ & $3.32\pm0.02$ & $13.43\pm0.01$ & 13795.1  & 12614 \\ \hline
$\sigma_\mathrm{int}=0.11$ & $0.305\pm0.008$ & $0.117\pm0.001$ & $3.30\pm0.02$ & $13.43\pm0.01$ & 12668.9  & 12614  \\ \hline
$\sigma_\mathrm{int}=0.12$ & $0.312\pm0.009$ & $0.116\pm0.001$ & $3.28\pm0.02$ & $13.44\pm0.01$ & 11663.3  & 12614 \\ \hline
\end{tabular}
\end{table*}

\section{Summary and discussion}\label{sec:summary}

We have developed a method for SN classification using their colors and parameters from a general function fit of their light curves, utilizing the Random Forest classification algorithm. Our method is independent of the SN models, and make no use of redshift information of the SN or its host galaxy. We have achieved performance comparable to other photometric classification methods. By varying the probability threshold we are able to obtain samples with different purity as needed. A sample with 99\% purity is chosen for our cosmological constraints study in this paper. 

We use the general function fit (``Bazin fit") to obtain accurate peak magnitudes and calculate peak colors, making our classification method independent of SN Ia light curve models. Although peak magnitudes and colors can be obtained using a SALT2-like SN Ia model, fitting non-Ia's to a Ia model would fail in most of the cases, and would result in a training sample with only a small number of non-Ia's if a fit probability cut is applied before classification, which may bias the machine learning classifier. Although the number of CC SNe in the training sample can be boosted by simulation, only the SNe simulated from very a few CC templates would survive the cut, so the training sample is still biased. Alternatively, we could use the SALT2 fit probability as a feature instead of making the cut. However, using features from a SALT2-like model would bring in correlations between the Ia light curve parameters (which are incorrect for CC SNe) and lead to biased features, while a general fitting function like the ``Bazin" function does not have such correlations built in and so may give a more unbiased feature set. Meanwhile, the ``Bazin fits" deliver other parameters such as $t_\mathrm{fall}$ and $t_\mathrm{rise}$, which are characteristics of the light curves, and are useful features for classification. 

We recognize that using the ``Bazin fit" would lose high SNR events, but those are only a small fraction and mainly low-redshift events that will very likely be followed up spectroscopically. In this paper we focus on the classification of SNe at higher redshifts, which are unlikely to be followed up spectroscopically. In the future, a combination of fitting methods using both SALT2 and the Bazin function could be used to include the low-z SNe.

We have obtained photo-z's for our photometric sample by fitting the SN light curves using the extended SALT2 model, with the nested sampling method. We show that initial conditions and parameter limits need to be set very carefully in order to obtain good results, especially when no host-galaxy prior is given. Our photometric redshifts have a mean bias ($\left<z_\mathrm{phot}-z_\mathrm{spec}\right>$) of 0.0120 with $\sigma\left( \frac{z_\mathrm{phot}-z_\mathrm{spec}}{1+z_\mathrm{spec}}\right) = 0.0294$ without using a host-galaxy photo-z prior, and a mean bias ($\left<z_\mathrm{phot}-z_\mathrm{spec}\right>$) of 0.0017 with $\sigma\left( \frac{z_\mathrm{phot}-z_\mathrm{spec}}{1+z_\mathrm{spec}}\right) = 0.0116$ using a host-galaxy photo-z prior.

We obtained a final photometric sample with further cuts on the photo-z errors and the light curve parameters from the SALT2 fit. Using our final photometric SN Ia sample derived without using host-galaxy photo-z prior, and assuming a flat $\Lambda$CDM model, we obtain a measurement of $\Omega_m$ of $0.305 \pm 0.008$ after bias correction, with statistical errors only and the intrinsic scatter set to $\sigma_\mathrm{int}= 0.11$. The fitted $\Omega_m$ is consistent with our simulations ($\Omega_m=0.3$). The fitted value varies a little with different choice of the intrinsic scatter terms. With the small statistical uncertainty due to the large sample size, the study of the systematic effects becomes more important. Here we aim at showing the capability of constraining cosmology using the photometric sample. We will leave the systematic studies for future work.     

\section*{Acknowledgements}
Some of the computing for this project was performed at the OU Supercomputing Center for Education \& Research (OSCER) at the University of Oklahoma (OU). Argonne National Laboratory’s work was supported under U.S. Department
of Energy, contract DE-AC02-06CH11357.

\appendix

\section{Quality cuts details}\label{appx1}
We present the numbers of SNe remaining after each quality cut in detail in Table \ref{table:quality_cut} and \ref{table:bazin_cut}.

\begin{table*}
\caption{Summary of number of SNe remaining for each type after each quality cut}
\begin{threeparttable}
\centering
\label{table:quality_cut}
\begin{tabular}{|c|c|c|c|}
\hline
                           &   Ia      &      II          &          Ibc   \\ \hline
Total number of SNe before any cuts      &   199400 (1)        & \multicolumn{2}{c|}{ 1941000 (1)}                 \\ \hline
Max SNR $> 5$ for 3 bands  &    62147 (0.31) & 67631 (0.035) & 14468 (0.007)             \\ \hline
\multirow{2}{*}{\parbox{5cm}{1 point before and 2 after the peak\tnote{a}$\;$ for 3 bands , 1 of which has max SNR$>5$}} & \multirow{2}{*}{48298 (0.24)} & \multirow{2}{*}{54900 (0.028)} & \multirow{2}{*}{11468 (0.006)} \\
                  &                   &                   &                   \\ \hline
Bazin fit success\tnote{b}$\;$ (all 6 bands) & 48159 (0.24) & 52342 (0.027) & 11311 (0.006) \\ \hline
Bazin parameter cuts\tnote{c}$\;$     & 26616 (0.13)  &  7960 (0.004)  & 4354 (0.002)  \\ \hline
Final fraction\tnote{d}$\;$           & 0.684  & 0.204  & 0.112 \\ \hline
\end{tabular}
\begin{tablenotes}
      {\footnotesize
      \item[a] Here “peak” refers to the highest flux point in the raw light curve whose SNR is greater than the median SNR of that band.
      \item[b] Here “success” refers to any fit that returns a set of values (does not return a “failure” by the curvefit program), whether they are in a reasonable range or not.
      \item[c] Detailed in Table \ref{table:bazin_cut}.
      \item[d] Fraction of types in the final sample (added up to 1).
      \item[e] Numbers in parentheses indicate the fractions.

      }  
\end{tablenotes}
\end{threeparttable}
\end{table*}

\begin{table*}
\caption{Number of SNe remaining for each type after each Bazin parameter cut}
\begin{threeparttable}
\centering
\label{table:bazin_cut}
\begin{tabular}{|c|c|c|c|}
\hline
 & Ia & II & Ibc \\ \hline
After pre-fit cuts & 48159 (1)&	52342 (1)&	11311 (1)   \\ \hline
\multirow{2}{*}{\parbox{4.2cm}{$t_{\mathrm{rise}} > 1$, and $t_{\mathrm{rise}}$ not close to 1 with tolerance = 0.01}} & \multirow{2}{*}{40337 (0.84)} & \multirow{2}{*}{34162 (0.65)} & \multirow{2}{*}{8320 (0.74)} \\
                  &                   &                   &                   \\ \hline
$-20<B<20$ & 38317 (0.80)&	17439 (0.33)&	6228 (0.55)  \\ \hline
$\chi^2/d.o.f < 10$ & 36041 (0.75)&	16308 (0.31)&	5995 (0.53) \\ \hline
$t_\mathrm{fall} < 150$ & 33580 (0.70)&	13038 (0.25)&	5742 (0.51)  \\ \hline
$t_{\mathrm{rise}} < t_{\mathrm{fall}}$ & 31248 (0.65)&	11201 (0.21)&	5219 (0.46) \\ \hline
$A < 5000$ & 31191 (0.65)&	11172 (0.21)&	5207 (0.46)  \\ \hline
$A_{\mathrm{err}} < 100$ & 27806 (0.58)&	9612 (0.18)&	4627 (0.41)  \\ \hline
$t_{0,\mathrm{err}} < 50$ &  27545 (0.57)&	9121 (0.17)&	4567 (0.40) \\ \hline
$t_{\mathrm{fall,err}} < 100$ & 26826 (0.56)&	8078 (0.15)&	4390 (0.39) \\ \hline
$t_{\mathrm{rise,err}} < 50$ & 26616 (0.55)&	7961 (0.15)&	4354 (0.38) \\ \hline
$A(Y), A(u) < 1000$ & 26616 (0.55)&	7960 (0.15)&	4354 (0.38) \\ \hline
\end{tabular}
\begin{tablenotes}
      {\footnotesize
      \item[$\dagger$] Numbers in parentheses indicate the fractions.

      }  
\end{tablenotes}
\end{threeparttable}

\end{table*}

\section{Details on the comparison between the peak magnitudes obtained using Bazin and SALT2}\label{appx2}

For a Bazin fit, the peak magnitudes are calculated as described in Section \ref{subsec:bazinfit}. For a SALT2 fit, the fitting delivers the SALT2 parameters, so that the model SED can be calculated using the fitted parameters and the peak magnitudes can be obtained by integrating the model SED. 
For each fitting method, we separate the sample into Ia and CC subsamples and calculate the mean and standard deviation of the difference between the obtained peak magnitudes and the true magnitudes (calculated using the simulated models) in 20 redshift bins, respectively. The results are shown in Figure \ref{fig:peakmags}.

For clarity, we refer to the differences between the SAMPLE peak magnitudes obtained using the MODEL and the true simulated peak magnitudes as ``the MODEL SAMPLE differences", where SAMPLE is replaced by Ia or CC, and MODEL is replaced by SALT2 or Bazin, respectively.
We find that the SALT2 Ia differences have the smallest scatter; the Bazin Ia differences have a slightly larger scatter, compared to the SALT2 Ia differences; and the CC differences for both SALT2 and Bazin have the largest and comparable scatters. Since the SALT2 model is known to work well on SNe Ia by its nature, the smallest scatter in the SALT2 Ia differences is expected. The slightly larger scatter in the Bazin Ia differences compared to the SALT2 Ia difference is also expected, since the fitting is performed band by band, without following the known Ia color relations. 

We notice that the mean values of the Ia differences follow similar trends, and are close to 0, indicating that there is no significant bias in the Ia peak magnitudes despite intrinsic scatters. The mean of the SALT2 CC differences show a larger deviation from 0, indicating some biases, although not significant given the large scatter. The mean of the Bazin CC differences is consistent with 0, except for the Y band, which shows a slightly larger deviation from 0. The larger deviation in the Y band may be due to the lower signal-to-noise-ratio (SNR).
For Bazin, the larger scatter in the CC differences could be due to the fact that the SNR is lower for the CC sample, compared to the Ia subsample. For SALT2, the observed larger scatter and slightly biased mean are the consequence of both a lower SNR and the use of an unsuitable model. While the SALT2 CC peak magnitudes appear to be not significantly biased, when looking at the resultant SALT2 parameters and plotting out the model light curves, we find that very often the parameters are near the edges of the parameter limits and shape of the model light curves are very uneven, indicating that the fit does not work well. Since at the edges of the parameter limits, the model uncertainty is larger, and the light curve itself is noisy and may have poor cadences, the fit and the quoted peak magnitudes are subject to larger uncertainties and are therefore less reliable. Meanwhile, when fitting the SALT2 model, no redshift information is given, so a CC SNe may be fit using the SALT2 model with wrong redshift. Thus the calculated peak magnitudes may be close to the truth, but the fit is internally wrong. If the host photo-z prior is applied, the resultant peak magnitudes may be even more biased.

Given the above reasons, we conclude that the Bazin function provides unbiased peak magnitudes for the purpose of this paper, and is more suitable for fitting the CC SNe. We will also explore better ways to obtain the peak colors in the future.

\begin{figure*}
\centering
\includegraphics[width=\textwidth]{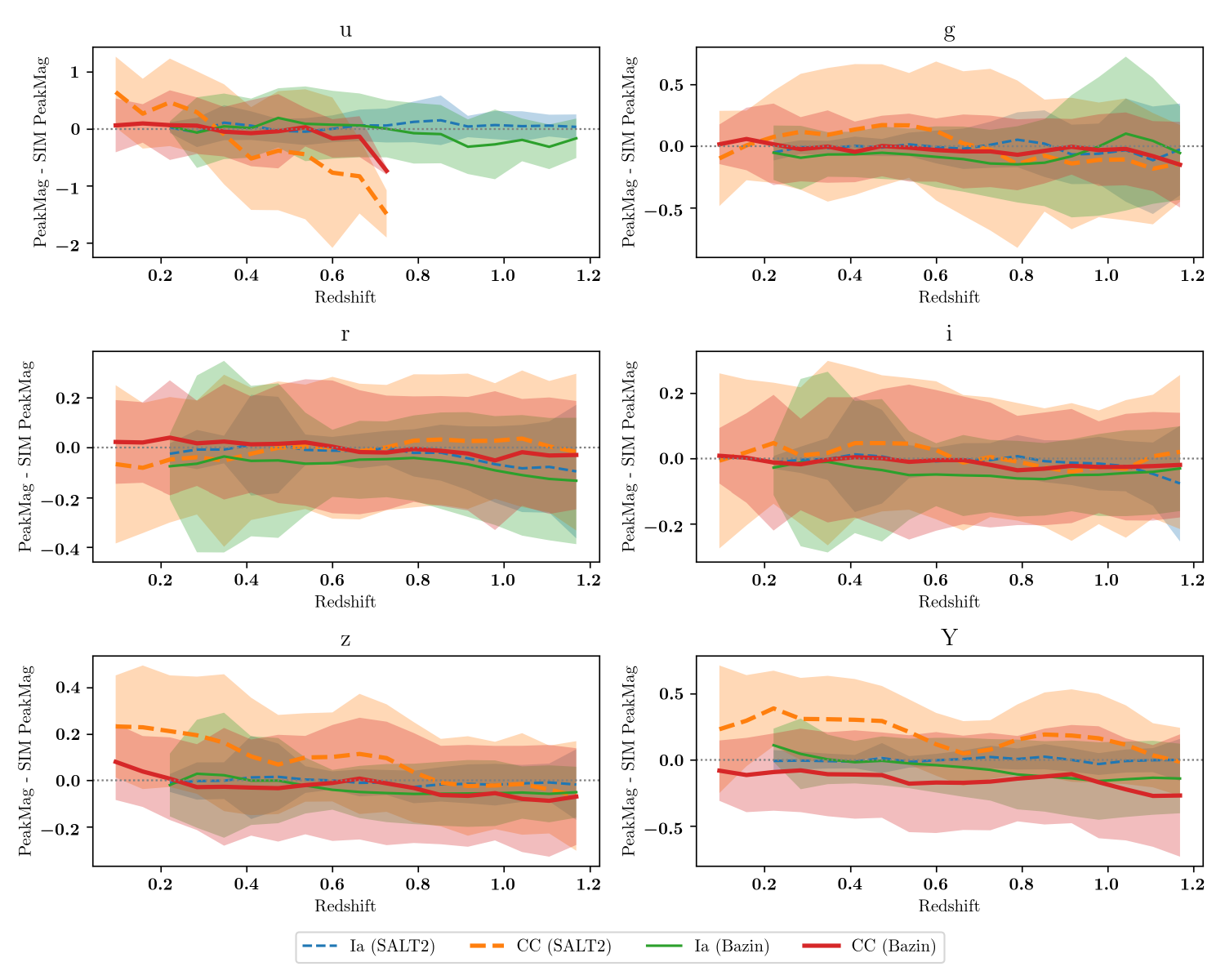}
\caption{Means and standard deviations of the peak magnitude differences. Lines are the means, thin and thick, dash and solid represents Ia and CC, SALT2 and Bazin respectively, shaded areas are the standard deviations.}
\label{fig:peakmags}
\end{figure*}

\end{document}